\documentstyle[12pt,thmsa,sw20aip]{article}

\input{tcilatex}
\begin{document}

\author{A. Tartaglia \\
Dip. Fisica, Politecnico, Corso Duca degli Abruzzi 24, I-10129 Torino, Italy.%
\\
E-mail: tartaglia@polito.it}
\title{Gravitational Aharonov-Bohm effect and gravitational lensing}
\maketitle

\begin{abstract}
Considering the spacetime around a rotating massif body it is seen that the
time of flight of a light ray is different whether it travels on one side of
the source or on the other. The difference is proportional to the angular
momentum of the body. In the case that a compact rapidly rotating object is
the source of a gravitational lensing effect, the contribution coming from
the above mentioned gravitational Aharonov-Bohm effect should be added to
the other causes of phase difference between light rays coming from
different images of the same object.

\end{abstract}

\section{Introduction}

The Aharonov-Bohm (AB) effect is due to the influence of the vector
potential on the phase of the wave function of a charged particle\cite{bohm}%
. The existence of a similar effect for gravitational fields also was
pointed out in connection with phenomena expected when working with rotating
reference frames or rotating sources of gravity. In fact many an author (see
for instance Aharonov and Carmi\cite{carmi}, Sakurai\cite{sakurai}, Semon%
\cite{semon}, Tsai and Neilson\cite{tsai}) showed how the Sagnac effect
could be explained in terms of an inertial AB effect, both at the classical
and at the quantum level. A truly gravitational effect was worked out by
Harris\cite{harris} exploiting the similarity between the electromagnetic
field and the description of the gravitational field in terms of
gravitoelectric and gravitomagnetic components to show precisely the
existence of an AB effect induced by the rotation of a massif body.

This gravitational contribution to what is otherwise known as the Sagnac
effect in a terrestrial environment (surface of the Earth or satellites
orbiting it) is very tiny\cite{tartaglia}, but there are situations of
astronomical interest where the conditions could be different.

One such field where to look for a gravitational AB effect is gravitational
lensing (GL). The influence of gravity on the path of a light ray was
considered long before the outset of general relativity, dating back to the
XVIII century; the idea of GL appeared as soon as in 1920 when it was
somehow foreshadowed by Eddington\cite{eddington}, then it slowly grew up in
theorists' consideration but it was only at the end of the 70's that an
observational evidence for such a phenomenon was found and since then the
examples have multiplied and with them the interest by the astronomers
community. Nowadays a rich literature exists exploring many details of the
lensing mechanism (a complete account of the theory may be found in \cite
{schneider}). One of the problems tackled when examining light coming from
different images of one single object is the one of the phase relations
between each of them and the others. A phase difference arises from the
different travel times of the light rays following different paths. The
difference in times has its origin both in the different length of the path
as seen by an inertial observer and in the intensity of the gravitational
fields encountered during the travel.

Now, coming back to the gravitational AB effect, one may expect that it also
contributes to the phase difference between light rays following different
paths, at least in the case that the bending source is a compact object
endowed with a strong gravitational field and a very high angular momentum.
The purpose of this paper is to evaluate this contribution. Section II
summarizes the essentials of the gravitational AB effect; section III
determines the gravitational AB contribution to the time delay, and finally
section IV contains the conclusions and a short discussion of the results.

\section{The gravitational Aharonov-Bohm effect}

The gravitational AB effect originates in the peculiarities of the space
time around a rotating mass. Rotation per se leads to a desynchronization of
clocks laid along a closed path contouring the rotation axis; this is the
actual explanation of the Sagnac effect, i.e. of the phase difference
between light rays co-rotating and counter-rotating with a turntable. This
happens because the worldlines of points on such turntable are helixes in
four dimensions and this implies that no unique space exists for them,
whereas time is polydromic\cite{anandan}.

A similar situation is induced by a Kerr metric, just as it is the case for
the polydromic scalar potential of the magnetic field of a current\cite
{harris}. Studying the Sagnac effect in a Kerr metric\cite{tartaglia} one
finds that a phase difference between light rays contouring the source of
the gravitational field in opposite directions is present even when the
light source (and observer) is not rotating. This is what most properly can
be qualified as a gravitational AB effect because it may be reconduced to a
situation where a stationary source sends light beams on opposite sides of a
rotating mass towards a stationary observer.

In the simple case of a stationary observer and a circular equatorial path
for the light beams, the difference in travel time between the two
oppositely rotating rays when they come back to the observer, is\cite
{tartaglia}: 
\begin{equation}
\delta \tau =\frac{4\pi }{c^2}\frac{R_S}R\frac a{\sqrt{1-\frac{R_S}R}}
\label{ritardo}
\end{equation}

Here $R_S$ is the Schwarzschild radius of the central body ($R_S=2GM/c^2$), $%
M$ is its mass, $a$ is the ratio between the angular momentum and the mass
and $R$ is the radius of the considered equatorial circumference.
Considering a light source diametrically opposite to the observer the time
difference between the two paths would of course be half the value of (\ref
{ritardo}).

\section{Phase difference originated by the gravitational Aharonov-Bohm
effect.}

The situation we consider is that of a light source and an observer
stationary and far away from each other. Somewhere in between there is a
massive and rotating object which partly deflects the light rays. The
configuration is the one shown in fig.1. For simplicity we assume that the
rotation axis of the (let us call it so) gravitational lens (GL) is
orthogonal to the plane of the figure.

\FRAME{ftbpFU}{2.9827in}{2.7951in}{0pt}{\Qcb{Scheme of the situation. $S$ is
the light source, $b$ is the bender or source of the gravitational field, $O$
is the observer. The external lines show the paths of the light rays; the
broken line is the trace of the plane of the gravitational lens.}}{}{%
ahar1.eps}{}

Space-time is almost everywhere flat; the curvature is relevant only in the
vicinity of the lens. Furthermore the distances $l_1$ and $l_2$ are supposed
to be far greater than the size of the region where the curvature matters.
As a result of this assumption and of the fact that in astronomical
situations the deviation angles are always very small, the lens is treated
as acting on a plane perpendicular to the line of sight from the observer.

This means that the light rays may be thought of as being straight lines
broken at the lens plane. To find out the position of the break point, where
the whole deviation is supposed to occur, the Fermat's principle can be
used. In practice the position of the break point must be such that the
proper (for the observer) arrival time of light be stationary under any
variation of the position itself\cite{schneider}.

Now, in our conditions, we assume that the metric around the lens is the
Kerr metric: 
\begin{equation}
g=\left( 
\begin{array}{cccc}
c^{2}\frac{r^{2}-R_{S}r+\rho ^{2}\cos ^{2}\vartheta }{r^{2}+\rho ^{2}\cos
^{2}\vartheta } & 0 & 0 & \rho c\frac{R_{S}r}{r^{2}+\rho ^{2}\cos
^{2}\vartheta }\sin ^{2}\vartheta \\ 
0 & -\frac{r^{2}+\rho ^{2}\cos ^{2}\vartheta }{r^{2}-R_{S}r+\rho ^{2}} & 0 & 
0 \\ 
0 & 0 & -\left( r^{2}+\rho ^{2}\cos ^{2}\vartheta \right) & 0 \\ 
\rho c\frac{R_{S}r}{r^{2}+\rho ^{2}\cos ^{2}\vartheta }\sin ^{2}\vartheta & 0
& 0 & -\frac{\left( \rho ^{4}+r^{2}\rho ^{2}\right) \cos ^{2}\vartheta +\rho
^{2}\left( r^{2}+R_{S}r\sin ^{2}\vartheta \right) +r^{4}}{r^{2}+\rho
^{2}\cos ^{2}\vartheta }\sin ^{2}\vartheta
\end{array}
\right)  \label{metrica}
\end{equation}

The origin of the coordinates is at the center of the lens and $\rho =a/c$.
The plane containing the light trajectories is so oriented that $a>0$
(counterclockwise rotation).

The coordinated time lapse along an infinitesimal null worldline is in
general (Latin indices span the space coordinates): 
\begin{equation}
dt=\frac 1{g_{00}}\left[ -g_{0i}dx^i\pm \sqrt{\left(
g_{0i}g_{0j}-g_{ij}g_{00}\right) dx^idx^j}\right]  \label{tempogen}
\end{equation}

Specializing to the Kerr metric and for $\vartheta =\frac \pi 2=$cost (\ref
{tempogen}) becomes 
\begin{equation}
dt_{\pm }=d\varphi \frac r{c\left( r-R_S\right) }\left[ -\rho \frac{R_S}r\pm
f\left( r,\varphi \right) \right]  \label{tempo}
\end{equation}

where $f\left( r,\varphi \right) =\sqrt{\frac{\rho ^{2}R_{S}^{2}}{r^{2}}%
+\left( 1-\frac{R_{S}}{r}\right) \frac{\rho ^{2}\left( r+R_{S}\right) +r^{3}%
}{r}+\left( 1-\frac{R_{S}}{r}\right) \frac{r^{2}}{r^{2}-R_{S}r+\rho ^{2}}%
\left( \frac{dr}{d\varphi }\right) ^{2}}$

Time, by default, flows toward the future ($dt>0$). To insure this we see
that when the ray goes past the lens on the left ($d\varphi >0$) it must be 
\begin{equation}
dt_{l}=d\varphi \frac{r}{c\left( r-R_{S}\right) }\left[ -\rho \frac{R_{S}}{r}%
+f\left( r,\varphi \right) \right]  \label{sinistra}
\end{equation}

When viceversa it passes on the right ($d\varphi <0$), it must be 
\begin{eqnarray}
dt_r &=&d\varphi \frac r{c\left( r-R_S\right) }\left[ -\rho \frac{R_S}%
r-f\left( r,\varphi \right) \right] =  \label{destra} \\
&&\left| d\varphi \right| \frac r{c\left( r-R_S\right) }\left[ \rho \frac{R_S%
}r+f\left( r,\varphi \right) \right]
\end{eqnarray}

The difference between (\ref{sinistra}) and (\ref{destra}) is the origin of
the time delay between the two beams. If, by any means, one could impose the
same geometrical path on both sides and if source, GL and observer are
alined, the total time difference is: 
\begin{equation}
\delta t=\frac{2}{c}\rho R_{S}\int_{0}^{\pi }\frac{d\varphi }{r-R_{S}}
\label{uguali}
\end{equation}

This situation corresponds geometrically to the formation of a Chwolson ring%
\cite{chwolson}\footnote{%
Usually called Einstein ring, but Einstein noticed the effect now named
after him only in 1936\cite{einstein}, whereas Chwolson did so in 1924.}. In
general one can expect that the angular momentum of the GL produces unequal
paths on the sides of the bender, if however the additional bending is small
with respect to the usual potential effect (\ref{uguali}) may be used to
estimate the contribution to the phase difference coming from the angular
momentum alone.

Accepting the idea that the light rays are straight, we can immediately
write their equation. The reference is made to fig 1; the result for $0\leq
\varphi \leq \pi /2$ is 
\begin{equation}
r_{1}=\frac{l_{1}\sin \alpha }{\sin \alpha \cos \varphi +\cos \alpha \sin
\varphi }  \label{erre1}
\end{equation}

and for ($\pi /2\leq \varphi \leq \varphi _s$) 
\begin{equation}
r_2=\frac{l_1\tan \alpha }{\sin \varphi -\frac{l_1}{l_2}\tan \alpha \cos
\varphi }  \label{erre2}
\end{equation}

Now we can introduce (\ref{erre1}) and (\ref{erre2}) into (\ref{uguali}) and
perform the integration. Considering the smallness of $\alpha $ the result
is (see appendix): 
\begin{equation}
\delta t\simeq 4\frac{\rho }{c}\frac{R_{S}}{l_{1}\alpha }=4\frac{a}{c^{2}}%
\frac{R_{S}}{l_{1}\alpha }=8\frac{G}{c^{4}}\frac{J}{l_{1}\alpha }
\label{finale}
\end{equation}

The corresponding phase difference for a radiation whose frequency is $\nu ,$
is of course: 
\begin{equation}
\delta \Phi =2\pi \nu \delta t=16\pi \nu \frac{G}{c^{4}}\frac{J}{l_{1}\alpha 
}  \label{fase}
\end{equation}

\section{Conclusion}

We have found a simple formula for the time delay (and phase difference)
between two light rays running on opposite sides of a rotating and
gravitating object. In the final formula, $\alpha $ is an observational
parameter which actually depends in turn on the mass and angular momentum of
the source. The determination of its value may be done, as said in sect.
III, using Fermat's principle, however for an order of magnitude estimate it
is enough to know that it is approximately 
\[
\alpha \simeq \frac{R_{s}}{b} 
\]

where $b$ is the impact parameter of the light beam. It is in turn 
\[
b\simeq l_{1}\alpha 
\]

then 
\[
\alpha \sim \sqrt{\frac{R_{S}}{l_{1}}}\sim \frac{1}{c}\sqrt{\frac{GM}{l_{1}}}
\]

Finally: 
\begin{equation}
\delta t\sim \frac{1}{c^{3}}\sqrt{\frac{G}{Ml_{1}}}J  \label{ordine}
\end{equation}

In our simplified situation $\delta t$ is the whole delay between the
arrivals of light rays from two different images of the same source. In a
more realistic situation one should add to (\ref{finale}) the geometrical
delay originated by the difference in length for the two paths and the
potential delay (here I am using the terminology found in \cite{schneider})
due to the fact that, if the paths are geometrically different, the light
rays cross regions with different gravitational potentials.

The actual value of the delay depends of course on the parameters of the
bender, but the order of magnitude formula (\ref{ordine}) makes one suspect
that it can be not entirely negligible. The conversion of the delay into a
phase shift discloses the possibility to evidence it by interferometry
techniques.

\section{Appendix}

In order to evaluate (\ref{uguali}) we start from the two expressions (\ref
{erre1}) and (\ref{erre2}).

Suppose $R_S<<l_1\sin \alpha .$ When $r\equiv r_1$ it is: 
\begin{eqnarray*}
\frac 1{r-R_S} &=&\frac 1{l_1\frac{\sin \alpha }{\sin \alpha \cos \varphi
+\cos \alpha \sin \varphi }-R_S} \\
&\simeq &\left( 1+\frac{R_S}{l_1\sin \alpha }\left( \sin \alpha \cos \varphi
+\cos \alpha \sin \varphi \right) \right) \frac{\left( \sin \alpha \cos
\varphi +\cos \alpha \sin \varphi \right) }{l_1\sin \alpha }
\end{eqnarray*}

Consequently the first part of the integral in (\ref{uguali}) is

\begin{eqnarray*}
I_1 &=&\frac 2c\rho R_S\int_0^{\frac \pi 2}\frac{d\varphi }{r-R_S} \\
&\simeq &\frac 1{2c}\rho \frac{R_S}{l_1\sin \alpha }\left( 4\sin \alpha +%
\frac{R_S}{l_1\sin \alpha }\left( \pi +4\sin \alpha \cos \alpha \right)
+4\cos \alpha \right) \\
&\simeq &\frac 2c\rho \frac{R_S}{l_1\sin \alpha }\left( \sin \alpha +\cos
\alpha \right)
\end{eqnarray*}

Now let us play the same game when $r\equiv r_2$. One has: 
\begin{eqnarray*}
\frac 1{r-R_S} &=&\frac 1{l_1\frac{\sin \alpha }{\cos \alpha \sin \varphi -%
\frac{l_1}{l_2}\sin \alpha \cos \varphi }-R_S} \\
&\simeq &\frac 1{l_1\sin \alpha }\left( \cos \alpha \sin \varphi -\frac{l_1}{%
l_2}\sin \alpha \cos \varphi \right) \left( 1+\frac{R_S}{l_1\sin \alpha }%
\left( \cos \alpha \sin \varphi -\frac{l_1}{l_2}\sin \alpha \cos \varphi
\right) \right)
\end{eqnarray*}

The second part of the integral in (\ref{uguali}) is: 
\begin{eqnarray*}
I_2 &=&2\frac \rho cR_S\int_{\pi /2}^\pi \frac{d\varphi }{r-R_S} \\
&\simeq &2\frac \rho c\frac{R_S}{l_1\sin \alpha }\left( \cos \alpha +\frac{%
l_1}{l_2}\sin \alpha \right)
\end{eqnarray*}

Finally the whole integral is: 
\begin{eqnarray*}
\delta t &=&I_1+I_2 \\
&\simeq &2\frac \rho c\frac{R_S}{l_1\sin \alpha }\left[ 2\cos \alpha +\left(
1+\frac{l_1}{l_2}\right) \sin \alpha \right]
\end{eqnarray*}

The angle $\alpha $ is necessarily small in an astronomical configuration,
then: 
\[
\delta t\simeq 4\frac \rho c\frac{R_S}{l_1\alpha } 
\]

\end{document}